\documentclass[aps,prb,twocolumn,showpacs,amsmath,amssymb,superscriptaddress,runinaddress]{revtex4}
\usepackage{graphicx}
\usepackage{dcolumn}
\usepackage{textcomp}
\usepackage{bm}
\usepackage[T1]{fontenc}

\newcommand{\bs}{$\mathbf{b}^{*}$}

\begin{document}

\preprint{\today}

\title{Writing and reading chiral domains in multiferroic DyMnO$_3$ using soft X-rays
}

\author{E. Schierle}
\affiliation{Helmholtz-Zentrum Berlin für Materialien und Energie, Wilhelm-Conrad-Röntgen-Campus BESSY II, Albert-Einstein-Str. 15, D-12489 Berlin, Germany}

\author{V. Soltwisch}
\affiliation{Helmholtz-Zentrum Berlin für Materialien und Energie, Wilhelm-Conrad-Röntgen-Campus BESSY II, Albert-Einstein-Str. 15, D-12489 Berlin, Germany}

\author{D. Schmitz}
\affiliation{Helmholtz-Zentrum Berlin für Materialien und Energie, Wilhelm-Conrad-Röntgen-Campus BESSY II, Albert-Einstein-Str. 15, D-12489 Berlin, Germany}

\author{R. Feyerherm}
\affiliation{Helmholtz-Zentrum Berlin für Materialien und Energie, Wilhelm-Conrad-Röntgen-Campus BESSY II, Albert-Einstein-Str. 15, D-12489 Berlin, Germany}

\author{A. Maljuk}
\affiliation{Helmholtz-Zentrum Berlin  für Materialien und Energie, Lise-Meitner-Campus, Hahn-Meitner-Platz~1, D-14109, Berlin, Germany}

\author{F. Yokaichiya}
\affiliation{Helmholtz-Zentrum Berlin  für Materialien und Energie, Lise-Meitner-Campus, Hahn-Meitner-Platz~1, D-14109, Berlin, Germany}

\author{D. N. Argyriou}
\email{argyriou@helmholtz-berlin.de}
\affiliation{Helmholtz-Zentrum Berlin  für Materialien und Energie, Lise-Meitner-Campus, Hahn-Meitner-Platz~1, D-14109, Berlin, Germany}

\author{E. Weschke}
\email{eugen.weschke@helmholtz-berlin.de}
\affiliation{Helmholtz-Zentrum Berlin für Materialien und Energie, Wilhelm-Conrad-Röntgen-Campus BESSY II, Albert-Einstein-Str. 15, D-12489 Berlin, Germany}

\date{\today}

\begin{abstract}
Sizeable $\mathbf b$ and $\mathbf c$ components of the 4f moments and pronounced circular dichroism  are observed in the ferroelectric phase of DyMnO$_3$ by soft X-ray diffraction at the Dy-$M_5$ resonance. This points to cycloidal order of the 4f moments and indicates that inversion-symmetry breaking in this material is not characteristic of the Mn spins alone. The circular dichroism allows to image chiral domains that are imprinted on the surface of a DyMnO$_3$ single crystal, exploiting the local charging by the X-ray beam via the photoelectric effect. These findings suggest a novel approach to control and image domains and domain walls in multiferroic materials.
\end{abstract}


\maketitle

The use of ferroelectrics or ferromagnets in device applications relies on the switching of the magnetization or electric polarization by external fields, a process largely involving the motion of domain walls. New proposals for high-density \emph{magnetic} data storage devices such as \emph{racetrack} memories directly utilize the domain walls that can be moved by pulses of electric current~\cite{parkin:04112008}. 
Multiferroics, on the other hand, offer the possibility to manipulate both electric and magnetic domains and domain walls by electric fields~\cite{zhao:823,cheong:13,tokunaga:558}.   
The structure of domains and domain walls in multiferroic oxides can be rather complex due to their dual ferroic nature, as found, e.g., in GdFeO$_{3}$ with ferroelectric, ferromagnetic and antiferromagnetic order~\cite{tokunaga:558}. 
Magnetic ferroelectrics like BiFeO$_3$ reveal a close connection between ferroelectric and antiferromagnetic domains, although magnetic order occurs 
at substantially lower temperatures than ferroelectricity in this material~\cite{zhao:823}.
In some of the perovskite rare-earth manganates of composition REMnO$_3$, on the other hand, ferroelectricity is intimately related to cycloidal magnetic order. The corresponding  chirality gives rise to circular dichroism in magnetic x-ray diffraction~\cite{fabrizi:237205}, however, direct imaging of these chiral domains has been elusive so far. 
The present work deals with chiral domains in    
DyMnO$_{3}$, a material that belongs to this class of improper ferroelectrics where spontaneous electric polarization $\mathbf{P}$ arises with the cycloidal ordering of the Mn spins (Fig.~1) \cite{kenzelmann:087206}. A microscopic mechanism explaining this phenomenon is provided by the antisymmetric Dzyaloshinski-Moriya interaction~\cite{sergienko:094434} 
that favours a non-collinear arrangement of spins. Such a magnetic order breaks the inversion symmetry of the lattice and provides for a natural coupling between magnetism and ferroelectricity, relating $\mathbf{P}$ directly to the magnetic structure~\cite{mostovoy:067601,PhysRevLett.95.057205,cheong:13}:

\begin{equation}
\mathbf{P} \propto \sum_{i,j}{\mathbf{e}_{i,j}\times (\mathbf{S}_{i}\times \mathbf{S}_{j})}.
\end{equation}

In this equation, $\mathbf{S}_{i}$ and $\mathbf{S}_{j}$ denote spins at the sites $i$ and $j$,
and $\mathbf{e}_{i,j}$ is the unit vector connecting the two sites. Accordingly, domains of opposite ferroelectric polarization are linked to magnetic cycloids of opposite chirality 
as sketched in Fig.~1. They can be discerned by the asymmetry in the scattering of left-hand and right-hand circularly polarized x-rays~\cite{fabrizi:237205}.

\begin{figure}[t]
\includegraphics* [scale= 0.4, trim= -30 0 -10 0, angle= -90] {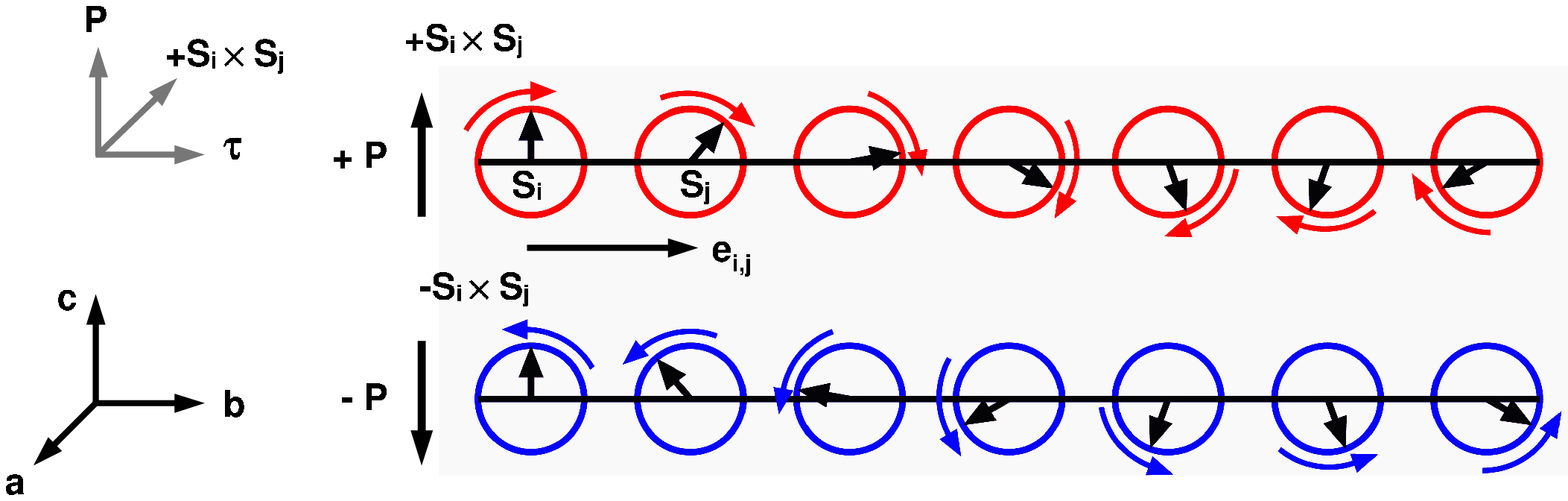}
\caption{\label{fig1} Relationship between the chirality of cycloidal spin order along the $\mathbf{b}$ axis and the direction of the ferroelectric polarization $\mathbf{P}$ along the $\mathbf{c}$ axis in DyMnO$_3$ according to Eq.~1. Domains with magnetic cycloids of opposite chirality correspond to opposite ferroelectric polarization.
  }
\end{figure}

DyMnO$_{3}$ displays a series of transitions upon cooling that are distinguished by both  electrical properties and magnetic order. Above $T_N \sim 39$~K, the material is  paramagnetic/paraelectric~\cite{kimura:224425}. Below this temperature, the Mn spins develop an ordered $\mathbf{b}$-axis magnetic moment that is sinusoidally modulated along the $\mathbf{b}$ axis with a propagation vector $\mathbf{\tau^{\mathrm Mn}}\sim0.39\cdot$\bs. This sinusoidal phase does not display ferroelectric polarization, which only occurs below
the ferroelectric transition at $T_{FE}\sim 18$~K, where the Mn spins adopt a cycloidal arrangement upon the development of 
an additional component of the Mn magnetic moment along the $\mathbf{c}$ axis~\cite{prokhnenko:057206}. This
cycloidal spin order within the $\mathbf{bc}$ plane of the perovskite unit cell with an axis of rotation of Mn spins around the $\mathbf{a}$ direction is illustrated in Fig.~1. The direction of the ferroelectric polarization is given directly by Eq.~1 to point along the $\mathbf{c}$ axis~\cite{kimura:224425}. 

The measurements were carried out at the UE46-PGM1 soft X-ray beamline of the
Helmholtz-Zentrum Berlin at BESSY II. 
Data were obtained with a UHV diffractometer designed at the Freie Universität 
Berlin~\cite{schierle:diss} using single crystals of DyMnO$_3$ as described in~\cite{feyerherm:134426}. Samples were cleaved perpendicular to the $\mathbf b$ axis,
in this way enabling scattering studies along the [0 $K$ 0] direction in specular geometry (Fig.~2a). Azimuth-dependent diffraction intensities could be recorded at low temperatures by an in-situ rotation of samples around the crystallographic $\mathbf b$ axis utilizing a sample holder developed at the University of Cologne~\cite{schlappa:diss}.   
Experiments were carried out on surfaces prepared under various conditions including cleavage in air as well as in the vacuum chamber, yielding consistent data from all samples. 
The scattered photons were detected without polarization analysis
using an Al-coated AXUV Si photodiode mounted behind a slit of rectangular shape. 

In the present study, we find that the magnetic order of the Dy-4f moments in DyMnO$_3$ tracks that of the Mn spins both in the sinusoidal and the cycloidal phase. 
We observe an intense (0~$\tau$~0) magnetic diffraction peak at the Dy-$M_5$ resonance (see Fig.~3c) that is closely connected to the Mn sinusoidal and cycloidal order, in particular $\tau = \tau^{\mathrm Mn}$~\cite{prokhnenko:057206}, which provides the basis for the results further discussed here. Polarization-dependent measurements of (0~$\tau$~0)
reveal that the $4f$ moments themselves exhibit cycloidal order, which has not been considered so far for this class of materials. 

The development of a $4f$ magnetic cycloid that is linked to 
the electric polarization of the material along $\mathbf{c}$ requires $\mathbf{b}$ and $\mathbf{c}$ components of the $4f$ magnetic moments. This was established by measuring the azimuthal dependence of the integrated intensity of the (0~$\tau$~0) magnetic diffraction peak in the sinusoidal compared to the cycloidal phase. The scattering geometry is shown in Fig.~2a, with the $\mathbf{b}$ axis in the scattering plane and perpendicular to the sample surface. 
Figure~2b displays the integrated intensity of (0~$\tau$~0) recorded with $\sigma$-polarized X-rays, $I_{\sigma}$, as a function of the azimuthal angle $\varphi$. The data were normalized to the intensity recorded with $\pi$-polarized X-rays, $I_{\pi}$, in order to account for possible sample inhomogeneities and plotted in a polar diagram.

\begin{figure}[t]
\includegraphics* [scale= 0.40, trim= -20 0 -10 0, angle= -90] {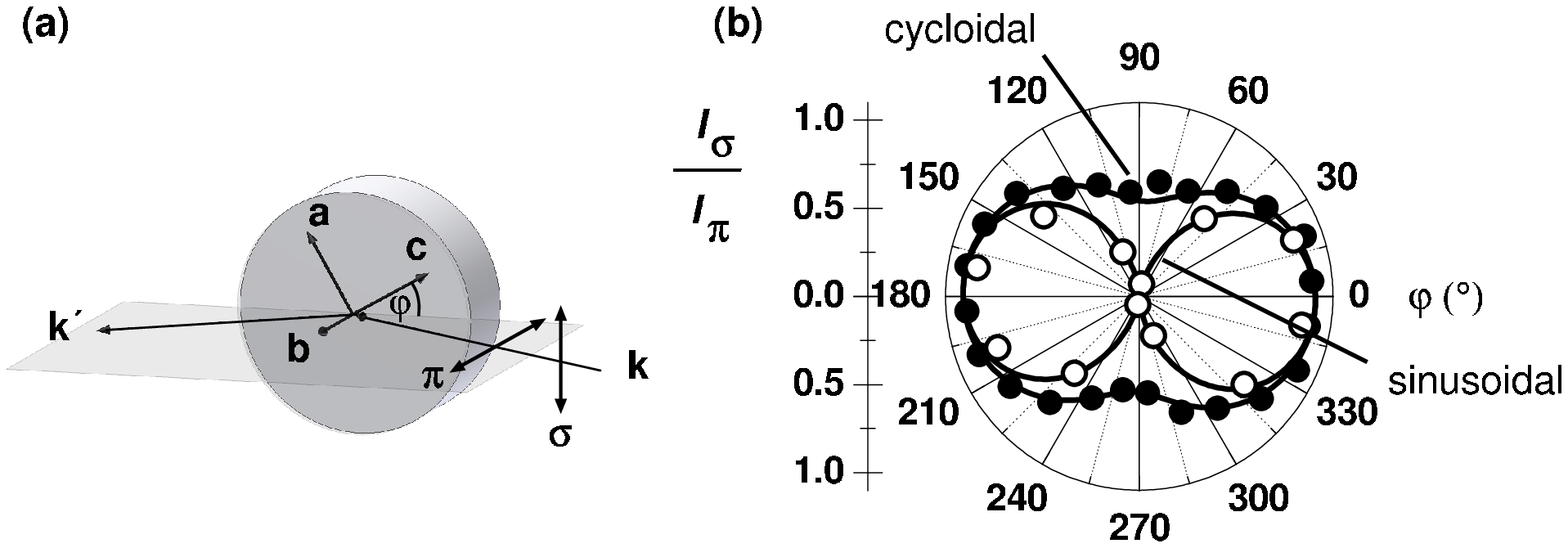}
\caption{\label{fig2} Azimuthal dependences of the (0~$\tau$~0) magnetic diffraction peak, recorded with linearly polarized X-rays ($\sigma$ and $\pi$). a) Scattering geometry showing the wavevectors of 
incident ($\mathbf k$) and scattered ($\mathbf k^\prime$) X-rays and their orientation
with respect to the crystal axes $\mathbf{a}$, $\mathbf{b}$, and $\mathbf{c}$. 
b) Azimuthal dependence of the integrated intensity of (0~$\tau$~0)
in the sinusoidal (open data points) and the cycloidal phase (solid data points).  
  }
\end{figure}

In the sinusoidal phase a node is observed for $\varphi=90^\circ$ and $\varphi=270^\circ$ (open circles). As can be inferred from the scattering geometry, 
these values correspond to a situation, where the $\mathbf{a}$ and the $\mathbf{b}$
axes  lie in the scattering plane. Since for resonant magnetic scattering incident $\sigma$-polarized X-rays are only sensitive to components of the magnetic moments in the scattering plane~\cite{hill:1996}, zero intensity means here that in the sinusoidal phase both, 
the $\mathbf a$ and the $\mathbf b$ component vanish, and the Dy $4f$ moment is characterized by a  $\mathbf{c}$-axis component only. In the cycloidal phase, the node is lifted (solid circles) and seizable intensity is observed at $\varphi=90^\circ$ and $\varphi=270^\circ$.
These data suggest a sinusoidally modulated $4f$ magnetic structure with the moments along $\mathbf c$ that develops into a cycloid with an additional $\mathbf b$ component - analogous to the Mn spins. Calculations using this scenario in fact result in a perfect description of the experimental data (solid lines through the data points in Fig.~2b). The additional $\mathbf b$ component of the $4f$ magnetic moment exhibits the same temperature dependence as found in hard x-ray resonant scattering~\cite{prokhnenko:057206}, resembling the behavior of an induced moment below $T_{FE}$. 

To confirm that the $\mathbf b$ and $\mathbf c$ components of the $4f$ magnetic moments indeed invoke cycloidal order, circularly polarized X-rays were employed, as the chirality of the magnetic cycloid would result in different scattering cross sections for left-hand and right-hand circular polarization, i.e. circular dichroism should be observed for a given magnetic domain~\cite{fabrizi:237205}. 
In the present multiferroic material, chirality and the direction of the ferroelectric polarization are closely related, and manipulating ($\mathbf{P}$) will immediately affect the chirality of the concomitant cycloid: If the polarization is changed from $+\mathbf{P}\rightarrow-\mathbf{P}$ this implies a change in the chirality of the spin cycloid from a clockwise rotation of the spin between ions $i$ and $j$ ($+\mathbf{P}$) to an anti-clockwise one ($-\mathbf{P}$) (Fig.~1) that results in a change of the sign of the dichroic signal.

\begin{figure}[t]
\includegraphics* [scale= 0.38, trim= -50 -5 -10 0, angle= -90] {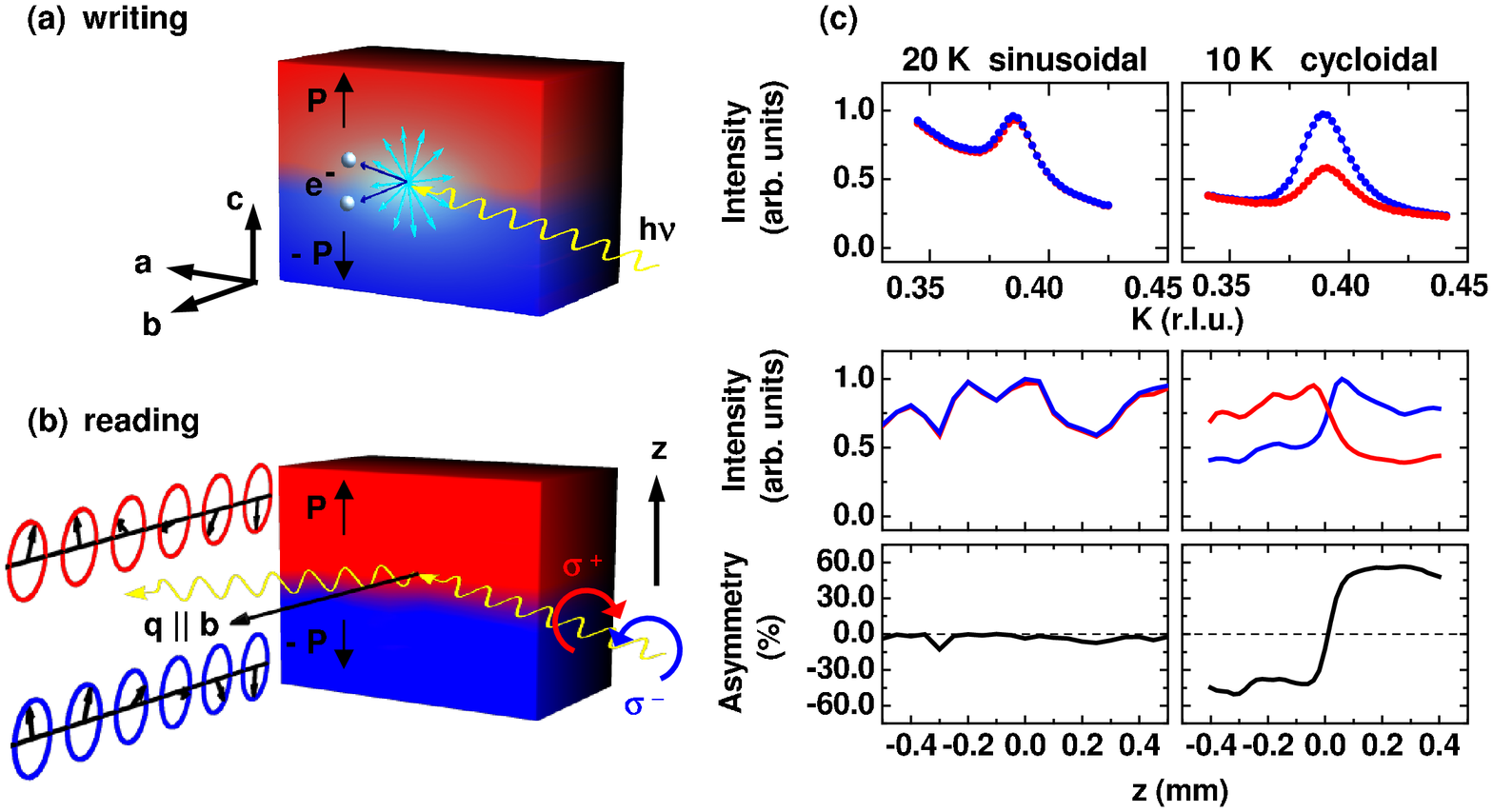}
\caption{\label{fig3} Multiferroic domains and X-ray circular dichroism
in DyMnO$_3$. (a) Two domains of opposite ferroelectric polarization are induced by the X-ray beam that creates a radial electric field by local charging via the photoelectric effect. (b) The concomitant cycloidal magnetic structures of opposite chirality yield X-ray circular dichroism in scattering that can be used to probe the domains. (c) (0~$\tau$~0) magnetic diffraction peaks recorded with right-hand (red) and left-hand (blue) circularly polarized X-rays in the sinusoidal paraelectric and the cycloidal multiferroic phase (top). The peak intensities as a function of the vertical sample position $z$ and the resulting asymmetries are shown on the bottom.
} 
\end{figure}

The measurement of circular dichroism requires domains of well-defined chirality in the area probed by the X-ray beam. In order to produce these domains in our sample, we have chosen a new approach that takes advantage of the local charging of the sample by the X-ray beam, rather than applying an external voltage across the crystal~\cite{fabrizi:237205}. 
As a ferroelectric oxide, DyMnO$_3$ is highly insulating and the emission of photoelectrons by the X-ray beam cannot be effectively compensated. In order to enhance the charging effect, a photon energy corresponding to the maximum of the $M_5$ resonance was chosen, where the photoionization cross section is resonantly enhanced.  In the present experiment, the effect of charging becomes substantial below $T \approx 100$~K, as observed by an almost vanishing drain current under the photon beam. Thus, due to the photoelectric effect, the photon beam induces an essentially radial electric field with a positively charged center at the position of the beam spot as illustrated in Fig.~3a. The field is confined to a surface layer of a few nm thickness corresponding to the escape depth of the photoelectrons~\cite{powell:1068}. In the absence of a magnetic field, the ferroelectric polarization in DyMnO$_3$ is constrained along the $\mathbf c$ axis~\cite{kimura:224425}, therefore, only the $\mathbf c$-axis component of this radial electric field will be effective for influencing the domains on the sample. We have utilized this mechanism to imprint two ferroelectric domains of opposite polarization on the sample surface by cooling into the ferroelectric phase below 18~K  with the beam illuminating a fixed spot on the sample ("writing"). This position sets the reference for the vertical coordinate at $z=0$. As the $\mathbf c$ axis of the DyMnO$_3$ crystal is vertical in the geometry of the experiment, a horizontal domain wall is produced that separates an upper domain of polarization $+\mathbf{P}$ (red) for $z>0$ from a lower one with $-\mathbf{P}$ (blue) for $z<0$. These two domains are characterized by magnetic cycloids of opposite chirality (Fig.~3b) that yield different scattering cross sections for right-hand ($\sigma^+$) and left-hand ($\sigma^-$) circularly polarized X-rays. 

In order to prove this scenario, we have measured the (0~$\tau$~0) magnetic diffraction peak with circularly polarized X-rays using a photon energy corresponding to the maximum of the Dy-$M_5$ resonance. The resonance permits fast recording of high-quality data, as shown for the paraelectric sinusoidal and the multiferroic cycloidal phases on the top panel of Fig.~3c. 
Obviously, right-hand (red) and left-hand (blue) circularly polarized X-rays yield substantially different intensities, notably, only in the cycloidal phase (right). In the sinusoidal phase that possesses no chirality, the circular dichroism is absent. In order to obtain consistent and reproducible results while measuring the dichroism, the photon flux had to be strongly reduced to $\approx 3$\% of the intensity applied for writing, otherwise, photoelectric charging effects can substantially alter the initial ferroelectric domain distribution during the measurement. Having established a strong circular dichroism of the (0~$\tau$~0) peak, we have measured the peak intensity as a function of the vertical sample position $z$ in the two phases. The raw data and the resulting asymmetry $A=(I(\sigma^+) - I(\sigma^-))/(I(\sigma^+) + I(\sigma^-))$, obtained after background subtraction, are shown in the lower panels of Fig.~3c. Evidently, $A$ is vanishing in the sinusoidal phase (left), while it is as large as 60\% in the multiferroic phase (right). 
The essential point here is that the asymmetry changes its sign at $z=0$, i.e., exactly at the position, where the X-ray beam was initially positioned, which is expected from the previous considerations. The writing process is not restricted to the particular sample position as we found that domains can be imprinted on the surface of the crystal at will. 

\begin{figure}[t]
\includegraphics* [scale= 0.4, trim= 0 0 0 -40] {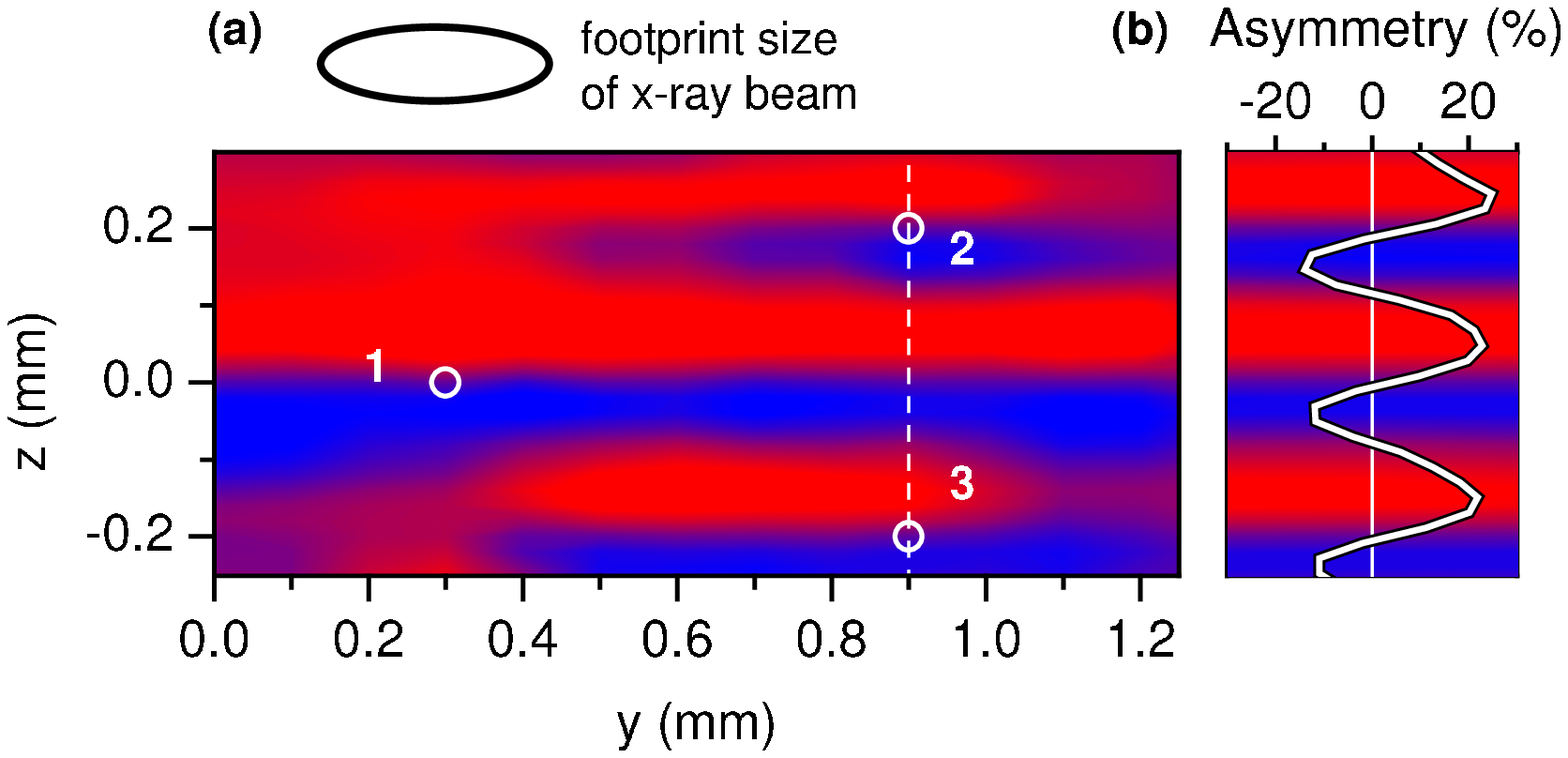}
\caption{\label{fig4} Two-dimensional pattern of chiral domains on the surface of a DyMnO$_3$ crystal in the cycloidal phase at T = 12 K with polarization $+\mathbf P$ (red) and $-\mathbf P$ (blue). The contrast is given by the circular dichroism asymmetry of the (0~$\tau$~0) magnetic diffraction peak intensity measured at the Dy-$M_5$ resonance; details are given in the text. A line scan as function of $z$ along the dashed white line is given in (b) and serves as an intensity scale for the asymmetry. The footprint size of the X-ray beam on the sample is given by the ellipse. 
}
\end{figure}

The possibility of writing and reading cycloidal magnetic domain structures is demonstrated for the case of a more complex pattern in Fig~4. Here, we show a two-dimensional 
domain pattern on the surface of the DyMnO$_3$ sample written by the local charging effect described above and recorded at 12~K. The figure displays the circular dichroic asymmetry as a function of sample position, coded in red for domains of $+\mathbf P$ and blue for $-\mathbf P$ (cf. Fig.~1). The intensity scale is given by the line scan in 
Fig.~4b, showing the asymmetry as a function of $z$ along the dashed white line. 

The initial patterning of the sample to produce the domain structure shown in Fig.~4
was accomplished by cooling the sample across $T_{FE}$ with the X-ray beam on the position labeled 1 in Fig.~4a. In this case, even a weak electric field component along $\mathbf c$ is sufficient to induce large domains of well-defined polarization. 
And indeed, two large domains of opposite chirality with long-range coherence along the $y$ direction for $z>0$ and $z<0$, respectively, are observed and can still be recognized in the final domain pattern. The subsequent writing procedures were carried out in a "softer" way in order not to destroy the initial patterning altogether. This was achieved by exposing the surface of the crystal to the X-ray beam at 12~K and then heating shortly up to 15~K, i.e., approaching but not crossing $T_{FE}$. The procedure was applied at the two spots labeled 2 and 3 in Fig.~4a, resulting in two additional domains. These domains are much more confined along both the $y$
and the $z$ direction than those obtained by cooling through the ferroelectric transition. In this way, chiral domains of variable size can be produced. Large domains on a mm scale are obtained by cooling through $T_{FE}$, while the soft writing process yields domains of strongly reduced size. Taking into account that the beam spot itself was of a considerable size of the order of $100 \times 300 \mu$m in the present experiment (Fig.~4a), even smaller domains may be envisaged using the technique. Apart from the beam spot size, domain sizes will also depend on the photon flux and the sample temperature that can be adapted to optimize the writing process. 
The asymmetry values of $\sim 20$\% displayed in Fig.~4 are smaller than those of Fig.~3 as the asymmetry was determined from the peak intensities alone without background correction in the former case. Nevertheless, it can be stated that the soft writing process yields the same maximum asymmetry as the initial patterning, showing that both processes can be equally effective, albeit on different length scales. 
The region, where the asymmetry changes sign is of the order of $100\mu$m along $z$ (Figs.~3c and 4b), which is essentially the vertical extension of the X-ray beam. Thus, an observation of domain walls and their internal structure is beyond the spatial resolution of the present experiment. It is to be noted, however, that in this class of materials, ferroelectric domain walls are expected to be of substantial thickness due to their magnetic nature~\cite{kagawa:057604}. Hence, they may be accessible by this technique using microfocus x-ray beams~\cite{rehbein:110801}. 
  
The present results show that ordering of the $4f$ magnetic moments in DyMnO$_3$ is strongly linked to that of the Mn-$3d$ spins, both in the sinusoidal and in the cycloidal phase. 
In particular, the substantial circular dichroism of the (0~$\tau$~0) peak associated with a magnetic cycloid parallel to the Mn spins reveals that inversion symmetry breaking in the multiferroic phase is not only accomplished by the Mn spins but also by the $4f$ moments, which has not been considered so far for this class of materials. 
Apart from these more fundamental aspects, 
the observation of the strong circular dichroism in soft x-ray scattering at the Dy-$M_5$ resonance provides a contrast mechanism to study chiral domains and domain walls. This resonance is characterized by a particularly large magnetic scattering cross section~\cite{ott:094412} and provides a very high sensitivity that will readily allow to study thin films and nanostructures, which is a straightforward extension of the rather surface sensitive method.
The present results also devise a novel approach to control multiferroic domain structures by photons. It uses the local charging of the sample surface by the photoelectric effect, which should generally apply to ferroelectrics as they are typically highly insulating materials. 
In combination with external electric fields, this might be a powerful tool to write and manipulate multiferroic domain walls and study their dynamics without the need of electrical currents. 
Further progress can be expected from the use of microfocus X-ray beams that are increasingly available at synchrotron radiation sources. This will expand the scope 
of the present experiments substantially towards manipulating and probing multiferroic domains with much higher lateral resolution, in this way accessing the essential length scales of these materials.


\end{document}